\documentclass[onecolumn, showpacs,%
  nofootinbib,aps,superscriptaddress,%
  prd,notitlepage,showkeys,10pt]{revtex4-1}

\usepackage{amssymb}
\usepackage{amsmath}
\usepackage{graphicx}
\usepackage{dcolumn}
\usepackage{hyperref}

\begin{document}

\title{Counterfactual quantum erasure: spooky action without entanglement}
\author{Hatim Salih}
\email[]{hatim.salih@bristol.ac.uk or salih.hatim@gmail.com}
\affiliation{School of Physics, HHWills Physics Laboratory, University of Bristol, Tyndall Avenue, Bristol BS8 1TL, United Kingdom}

%\author{ Boubacar Kante}
%\affiliation{Department of Electrical and Computer Engineering, University of California San Diego, La Jolla, California 92093-0407, USA}

\date{\today}

\begin{abstract}
We combine the eyebrow-raising quantum phenomena of erasure and counterfactuality for the first time, proposing a simple yet unusual quantum eraser: A distant Bob can decide to erase which-path information from Alice's photon, dramatically restoring interference---without previously-shared entanglement, and without Alice's photon ever leaving her lab.
\end{abstract}


\maketitle

Quantum erasure was first proposed by Scully and Druhl more than three decades ago \cite{Scully and Druhl}, sending shockwaves through the physics community. While early debates on double-slit interference going back to Bohr and Einstein \cite{Bohr} focused on Heisenberg's uncertainty principle as preventing one from learning which slit a particle went through while at the same time observing interference, quantum erasure put the focus on entanglement instead, a concept brought to light by Einstein and colleagues in the EPR paper \cite{EPR}. Scully and Druhl showed that it was possible to place a which-path tag on individual particles passing through a double-slit interferometer without disturbing them, thus throwing the uncertainty principle out of the discussion. Interference, however, is still lost because entanglement provides which-path information. The mere possibility of obtaining such information, regardless of whether it is actually obtained or not, is enough to destroy interference. Erasing which-path information, even long after the particles have been detected, remakably restores interference, seemingly allowing one to alter the past \cite{Kim, Walborn, Ahranov}. What is actually altered, however, is what one can say about the past---an argument for Bohr's view of physics as not being about what the world {\it is}, but rather what can be {\it said} about the world.

Practically, quantum erasure has been used to entangle, for the first time, two different-colour photons \cite{Zao}, and more recently, to propose a new protocol for quantum key distribution (QKD) that promises inherent security against powerful detector-targetting attacks \cite{Salih2016a}.

Counterfactuality, on the other hand, gleans information from events that could have happened but did not in fact take place. But information is physical---it is always manifested in physical form. The basic idea behind our present scheme is that information counterfactually communicated from Bob to Alice---that is without any particles travelling between them---can be made to manifest itself as a flip in the polarisation of Alice's photon. This allows us to combine the two phenomena of erasure and counterfactuality, proposing a simple yet unusual quantum eraser.

\begin{figure}
\centering
\includegraphics[width=0.5\textwidth]{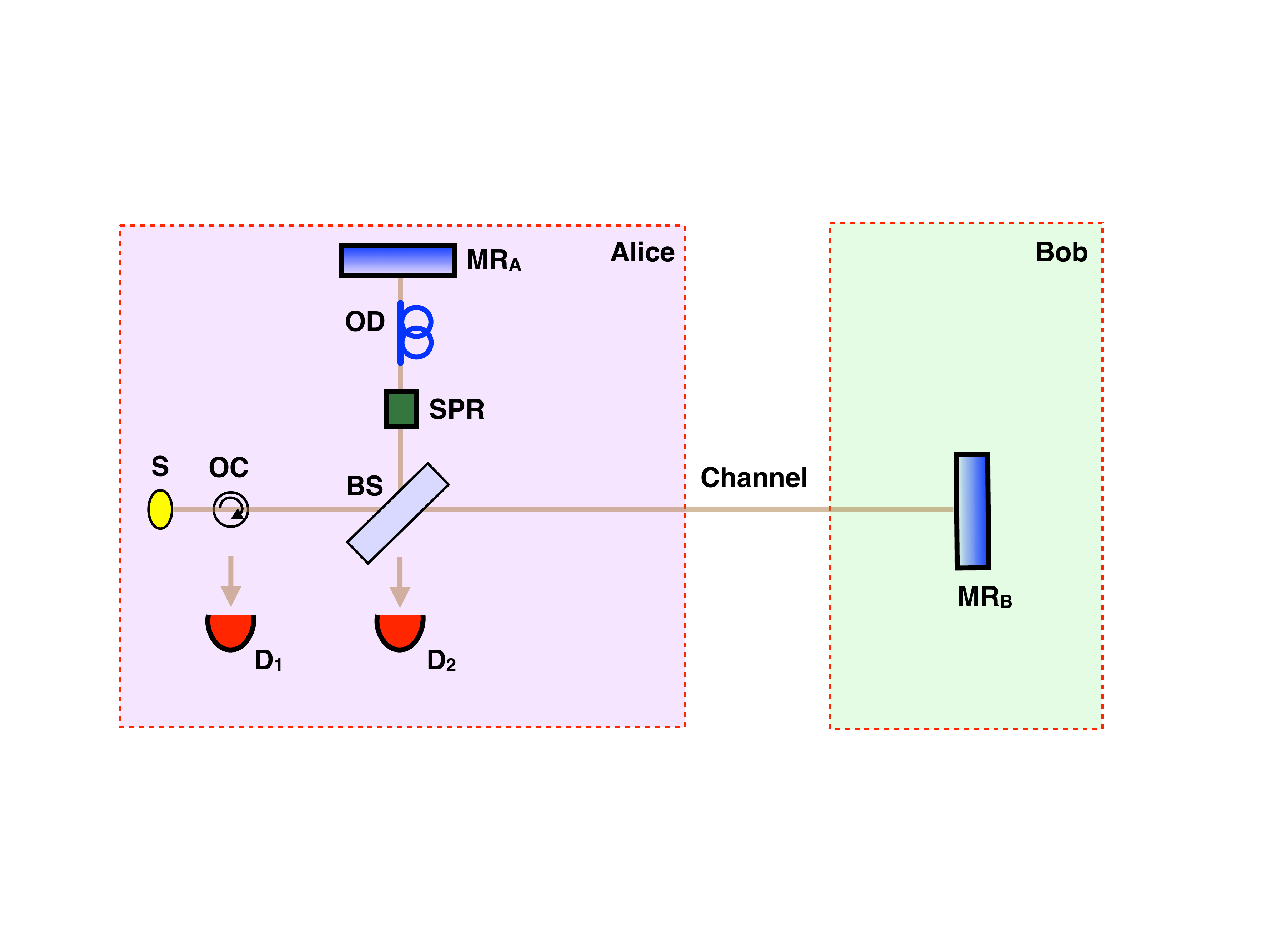}
\caption{\label{FIG1} Which-path information destroys interference. Single-photon source $S$ emits an $H$-photon towards the right. In this Michelson setup, interference of the photon components reflected off mirrors $MR_A$ and $MR_B$ means that detector $D_2$ always clicks. Optical delay $OD$ ensures that effective path-lengths match. Optical circulator $OC$ directs any photon coming from the right towards $D_1$. Flipping the polarisation of the photon component reflected by $MR_A$ towards $BS$, by means of switchable polarisation rotator $SPR$, provides a which-path tag. Interference is then destroyed. Detectors $D_1$ and $D_2$ are now equally likely to click.}
\end{figure}

Let's start with the Michelson interferometer of FIG. \ref{FIG1}. Alice's horizontally polarised $H$-photon, emitted by single-photon source $S$, encounters beam-splitter $BS$, which puts it in an equal superposition of traveling upwards towards mirror $MR_A$, and traveling to the right towards Bob's mirror, $MR_B$. These two components are reflected by mirrors $MR_A$ and $MR_B$ back to $BS$. By means of switchable polarisation rotator $SPR$, the polarisation of the part of the superposition incident on $BS$ from above can be flipped to $V$. There are two scenarios. First, if the polarisation of the part of the superposition incident on $BS$ from above is not flipped, by not applying $SPR$, no which-path information is available. Interference takes place, with detector $D_2$ always clicking. Second, if the polarisation of the part of the superposition incident on $BS$ from above is flipped to $V$ by applying $SPR$ appropriately, which-path information is imprinted. Interference does not takes place, with detectors $D_1$ and $D_2$ equally likely to click.

\begin{figure}
\centering
\includegraphics[width=0.5\textwidth]{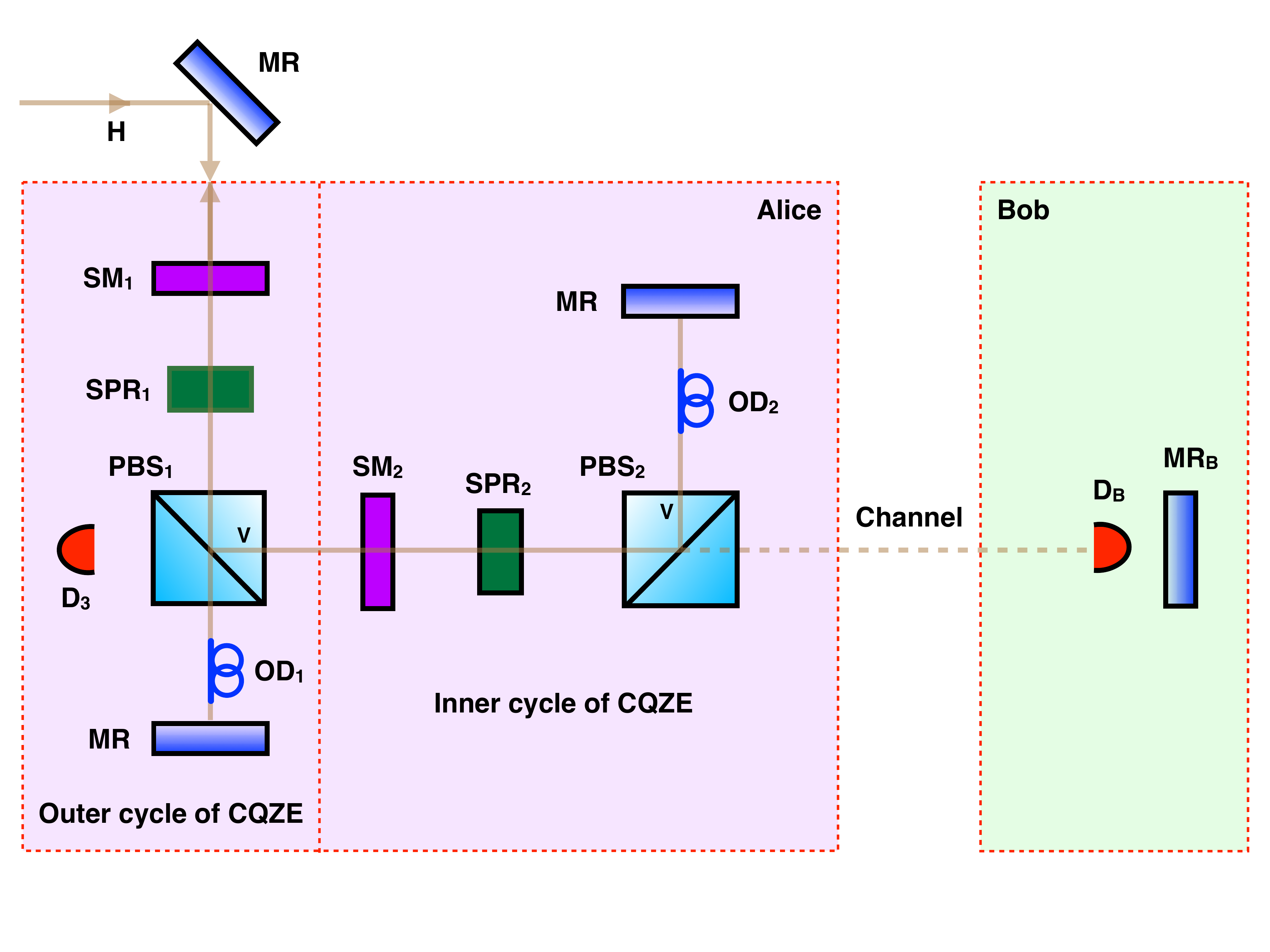}
\caption{\label{CQZE}The working of the Chained Quantum Zeno Effect (CQZE) for the case of Bob choosing to block the channel. We illustrate the operation of the CQZE using the minimum number of outer cycles, two. To start with, switchable mirror $SM_1$ is switched off letting Alice's $H$ photon in before it is switched on again. Using switchable polarisation rotator $SPR_1$ the following rotation is applied to the photon, $\left| \text{H} \right\rangle \to 1/\sqrt2(\left| \text{H} \right\rangle + \left| \text{V} \right\rangle)$, before it is switched off for the rest of this outer cycle. The $V$ part of the superposition is reflected towards Bob using polarising beamsplitter $PBS_1$. Switchable mirror $SM_2$ is then switched off to let the $V$ part of the superposition into the inner interferometer before it is switched on again. Using switchable polarisation rotator $SPR_2$, the following rotation, $\left| \text{V} \right\rangle \to \cos \frac{\pi}{2N} \left| \text{V} \right\rangle - \sin \frac{\pi}{2N} \left| \text{H} \right\rangle$, is then applied before it is switched off for the rest of this inner cycle. Polarising beamsplitter $PBS_2$ passes the $H$ part of the superposition towards Bob while reflecting the $V$ part. By blocking the channel, Bob effectively makes a measurement. Unless the photon is lost to $D_B$, the part of the photon superposition inside the inner interferometer ends up in the state $\left| \text{V} \right\rangle$. The same applies for the next $N-1$ inner cycles. Switchable mirror $SM_2$ is then switched off to let this part of the superposition, whose state has remained $\left| \text{V} \right\rangle$, out. In the next outer cycle, $SPR_1$ is switched on to rotate the photon's polarisation from $1/\sqrt2(\left| \text{H} \right\rangle + \left| \text{V} \right\rangle)$, assuming large N, to $\left| \text{V} \right\rangle$, before it is switched off for the rest of the final outer cycle. $PBS_1$ reflects the photon towards Bob. As before, after N inner cycles, provided it is not lost to $D_B$, the photon remains in the state $\left| \text{V} \right\rangle$. Finally, $SM_1$ is switched off to allow the photon, whose final state is $\left| \text{V} \right\rangle$, out. (Note that For the case of Bob not blocking the channel, it can be shown that repeated measurement by detector $D_3$ means that Alice's exiting photon is $H$-polarised in the end.) Optical delays $OD$ ensure that effective path-lengths match. $MR$'s are mirrors.}
\end{figure}

\begin{figure}
\centering
\includegraphics[width=0.5\textwidth]{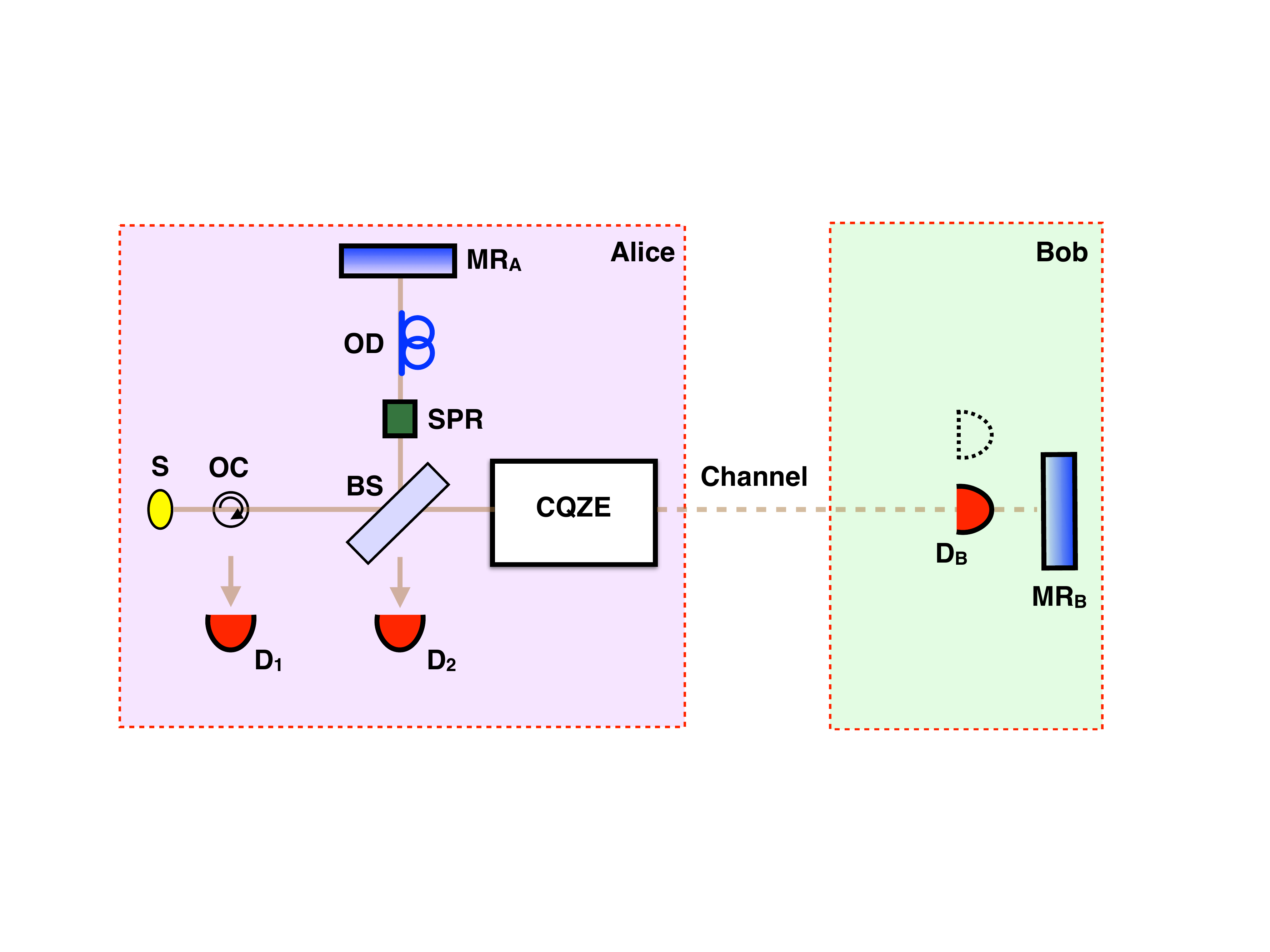}
\caption{\label{FIG2}Counterfactual erasure. Single-photon source $S$ emits an $H$-photon towards the right. Using the chained quantum Zeno effect (CQZE) module, the which-path tag imprinted by $SPR$ can be erased. Choosing to block the channel, Bob counterfactually erases which-path information by flipping the polarisation of the photon component travelling horizontally towards him. We can be sure that the photon has not traversed the channel, otherwise $D_B$ would have clicked. Interference is recovered, with $D_2$ virtually always clicking for large enough number of CQZE cycles. On the other hand, if Bob chooses not to block the channel, which-path information is not erased, $D_1$ and $D_2$ are equally likely to click, and interference is not recovered. In other words, Bob can remotely decide whether Alice observes interference or not without the photon ever leaving her station.}
\end{figure}

We now unveil counterfactual erasure. Using the chained quantum Zeno effect (CQZE) \cite{Hosten, Noh, Salih, Salih2014a, Salih2014b}, whose inner working is explained in FIG. \ref{CQZE}, and which has recently been experimentally demonstrated \cite{Cao}, Bob can decide to remotely flip the polarisation of the part of the photon superposition travelling from $BS$ towards $MR_B$ in FIG. \ref{FIG2} by merely blocking the channel, without Alice's photon leaving her station. 

Crucially, which-path information can be completely erased, thus restoring complete destructive interference at $D_1$. In the limit of a large number of inner cycles $N$ and outer cycles $M$ (with the number of outer cycles $M \ll N$) and given ideal implementation, detector $D_2$ always clicks.

Note that had Bob chosen not to block the channel, the polarisation of the part of the photon superposition travelling towards $MR_B$ would not have been flipped by CQZE \cite{Salih}. Erasure of which-path information would not have taken place and interference would not have been restored, with $D_1$ and $D_2$ equally likely to click.

The CQZE relies on two quantum phenomena, interaction-free measurement \cite{Elitzur, Kwiat1} and the quantum Zeno effect \cite{Sudarshan, Kwiat2}. In interaction-free measurement the mere presence of an obstructing object inside an interferometer destroys interference, allowing the object's presence to sometimes be inferred without interacting with any particle. The quantum Zeno effect on the other hand refers to the fact that repeated measurement of an evolving quantum object inhibits its evolution, an effect that brings to mind the proverbial watched kettle that does not boil. The quantum Zeno effect is used here to push the efficiency of interaction-free measurement towards unity.

The counterfactuality of the CQZE is based on the fact that any photon going into the channel is necessarily lost, which means that photons detected by Alice at $D_1$ or $D_2$ could not have travelled to Bob. From FIG. \ref{FIG2}, counterfactuality is ensured for the case of Bob blocking the channel: had the photon gone into the channel, detector $D_B$ would have clicked. For the case of Bob not blocking the channel, had the photon gone into the channel, detector $D_3$ would have clicked. Counterfactuality for the case of Bob not blocking the channel, which was disputed \cite{Vaidman, Salih reply, Vaidman2015}, has recently been proven using a consistent histories approach \cite{Salih2016b}.

%The fact that quantum erasure is brought about by Bob choosing to block the channel may at first sight suggest that the chaining of the Zeno effect, in other words having the outer cycles, is not needed---since the outer cycles were introduced in the first place only to ensure counterfactuality for the case of Bob not blocking. The case of Bob blocking is already counterfactual. The problem however is that without the outer cycles the scenario is reversed: polarisation is not flipped for Bob blocking. The chained quantum Zeno effect is therefore necessary.

The CQZE employs $N$ inner cycles nested within $M$ outer cycles. While, as can be inferred from the explanation in the caption of FIG. \ref{CQZE}, a smaller number of outer cycles does not lead to more output errors, a smaller number of inner cycles does lead to more output errors for the case of Bob blocking. The larger $N$ is, the closer to $V$ the polarisation of the part of the photon superposition travelling towards $MR_B$ is rotated, the more perfect the erasure. For a given $M$ and $N$, for the case of Bob blocking, the error can be obtained from the following recursion relations,

\begin{equation}
X[m] = \cos (\frac{\pi }{2M})X[m - 1] - \sin (\frac{\pi }{2M})Y[m - 1].
\end{equation}  

\begin{equation}
Y[m] = \left ( \sin (\frac{\pi }{2M})X[m - 1] + \cos (\frac{\pi }{2M})Y[m - 1] \right ) \cos^N (\frac{\pi }{2N}).
\end{equation}   

where $m$ corresponds to the end of the the m-th outer cycle, $X[M]$ and $Y[M]$ are the unnormalised probability amplitudes for the $H$ and $V$ components exiting the CQZE respectively. $X[M]$ is therefore the error term causing detector $D_1$ to incorrectly click. It approaches zero for large $N$,  

The quality of erasure can be measured by interference visibility, defined as $\frac{I_{max}-I_{min}}{I_{max}+I_{min}}$, where $I_{max}$ and $I_{min}$ are light intensities at detectors $D_2$ and $D_1$ respectively. $I_{min}$ and $I_{max}$ are proportional to the squared moduli of the probability amplitudes summed at detectors $D_1$ and $D_2$ respectively. By the action of $BS$ on the components $\frac{1}{\sqrt{2}}\left| \text{V} \right\rangle$, reflected off $MR_A$, and $\frac{1}{\sqrt{2}}(X[M]\left| \text{H} \right\rangle + Y[M]\left| \text{V} \right\rangle)$, exiting CQZE, we get,

\begin{equation}
Visibility = \frac{2Y[M]}{(X[M])^2 + (Y[M])^2 + 1}.
\end{equation} 

\begin{figure}
\centering
\includegraphics[width=0.5\textwidth]{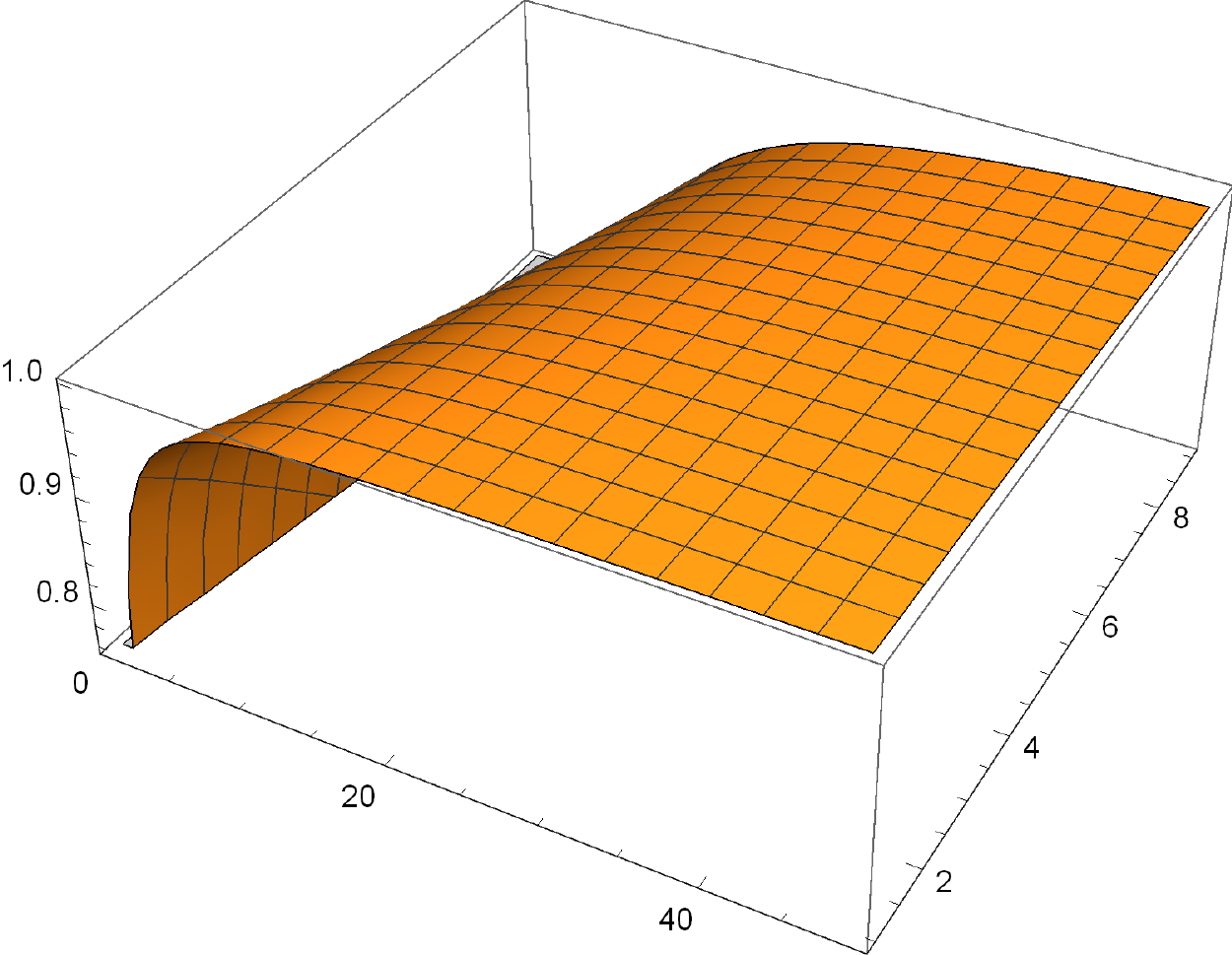}
\caption{\label{FIG3}Interference visibility of counterfactual erasure for number of outer cycles $M$ up to 10, and number of inner cycles $N$ up to 50. Visibility approaches unity for large $N$. Ideal implementation is assumed.}
\end{figure}

For instance, assuming ideal implementation, for a number of outer and inner cycles, $M$=2 and $N$=4, interference visibility is already 89$\%$. While for $M$=2 and $N$=14, interference visibility is 99$\%$. FIG. \ref{FIG3} plots interference visibility for $M$ up to 10 and $N$ up to 50. We note that all elements of this scheme are implementable using current technology.

Einstein, one imagines, would have been surprised, to put it mildly, by Scully and Druhl's quantum eraser. One wonders what he might have thought of the spooky-action-without-entanglement presented here---where we have shown how a distant Bob can choose to erase which-path information from Alice's photon counterfactually, that is without it ever leaving her lab, dramatically restoring interference.

\section*{ethics} 
There are no ethical issues.

\section*{data accessibility}
Not applicable.

\section*{author contributions} 
I am the sole author of this work.

\section*{competing interests} 
I have no competing interests.

\section*{funding} 
This work is funded by Qubet Research, a UK start-up in quantum technology.

\begin{acknowledgments}
I thank Sam L. Braunstein and Yuan Cao for useful comments. 

\end{acknowledgments}

%\section*{Appendix} 

\clearpage

\end{document}